\documentclass[%
 reprint,
 amsmath,amssymb,
pre,
]{revtex4-2}

\usepackage{graphicx}% Include figure files
\usepackage{dcolumn}% Align table columns on decimal point
\usepackage{bm}% bold math
\usepackage{hyperref}% add hypertext capabilities
\usepackage[utf8]{inputenc} % allow utf-8 input
\usepackage{url}            % simple URL typesetting
\usepackage{booktabs}       % professional-quality tables
\usepackage{amsfonts}       % blackboard math symbols
\usepackage{nicefrac}       % compact symbols for \frac{1}{2}, etc.
\usepackage{microtype}      % microtypography
\usepackage{lipsum}
\usepackage{graphicx}
\usepackage{amsmath}
\usepackage{cleveref}
\usepackage{tikz}
\usetikzlibrary{calc,arrows}
\usepackage[colorinlistoftodos,prependcaption]{todonotes}
\usepackage[mathlines]{lineno}% Enable numbering of text and display math
\begin{document}

\title{Exact Analytical Solution of the One-Dimensional Time-Dependent Radiative Transfer Equation with Linear Scattering}

\author{Vladimir Allaxwerdian}
\email{allaxwerdian@yandex.ru}
\author{Dmitry~V.~Naumov}
\affiliation{%
JINR, Dubna, Russia 141980
}
\email{dnaumov@jinr.ru}
\date{\today}% It is always \today, today,
\begin{abstract}
The radiative transfer equation (RTE) is a cornerstone for describing the propagation of electromagnetic radiation in a medium, with applications spanning atmospheric science, astrophysics, remote sensing, and biomedical optics. Despite its importance, an exact analytical solution to the RTE has remained elusive, necessitating the use of numerical approximations such as Monte Carlo, discrete ordinate, and spherical harmonics methods.

In this paper, we present an exact solution to the one-dimensional time-dependent RTE. We delve into the moments of the photon distribution, providing a clear view of the transition to the diffusion regime. This analysis offers a deeper understanding of light propagation in the medium.

Furthermore, we demonstrate that the one-dimensional RTE is equivalent to the Klein-Gordon equation with an imaginary mass term determined by the inverse reduced scattering length. Contrary to naive expectations of superluminal solutions, we find that our solution is strictly causal under appropriate boundary conditions, determined by the light transport problem.

We validate the found solution using Monte Carlo simulations and benchmark the performance of the latter. Our analysis reveals that even for highly forward scattering, dozens of random light scatterings are required for an accurate estimate, underscoring the complexity of the problem.

Moreover, we propose a method for faster convergence by adjusting the parameters of Monte Carlo sampling. We show that a Monte Carlo method sampling photon scatterings with input parameters $(\mu_s,g)$, where $\mu_s$ is the inverse scattering length and $g$ is the scattering anisotropy parameter, is equivalent to that with $(\mu_s(1-g)/2,-1)$. This equivalence leads to a significantly faster convergence to the exact solution, offering a substantial improvement of the Monte Carlo method for the one-dimensional RTE.

Our findings not only contribute to the theoretical understanding of the RTE but also have potential implications for improving the numerical methods used to approximate it.
\end{abstract}

\keywords{One-dimensional time-dependent RTE,faster Monte Carlo convergence,exact solution}%Use showkeys class option if keyword
\maketitle

%\tableofcontents
\section{Introduction}
The study of light propagation in random media is critical to many fields in fundamental and applied science. The Radiative Transfer Equation (RTE), a classical integro-differential equation, is a useful approximation for describing light propagation in these media. While the one-dimensional RTE is applied to various physical systems, such as atmospheric science~\cite{bohren2006fundamentals,lenoble1985radiative}, astrophysics~\cite{Rybicki1985}, remote sensing~\cite{tsang1978radiative}, and biomedical optics~\cite{wang2007biomedical}, an exact analytic solution is not available for even simple cases. Although a solution exists for the steady-state one-dimensional RTE~\cite{Adamson1975}, there is no known solution for the time-dependent one-dimensional RTE.

The main goal of this study is to derive an analytic solution to the time-dependent one-dimensional RTE for the propagation of light in random media, such as water or ice. Our objective is to use this solution and its generalization to the three-dimensional case to simulate the response of neutrino telescopes~\cite{Aartsen:2016nxy,Adrian-Martinez:2016fdl,AlvarezSanchez:2019gns,Avrorin:2021ogq}.

This article presents the exact solution to the one-dimensional RTE. The paper is structured as follows. \Cref{sec:the_solution} formulates the one-dimensional RTE and its solution. In~\cref{sec:discussion}, we discuss the results. In particular, in~\cref{sec:discussion_moments} the moments of the photon's distribution are discussed which allows to explicitly observe a change to the diffusion regime. In~\cref{sec:tachyon} we observe that one-dimensional RTE is equivalent to the Klein-Gordon equation with the imaginary mass term determined by the inverse reduced scattering length. However, in a stark contrast to naive expectations of superluminal solution,  the solution which we find is strictly casual under appropriate boundary conditions. In~\cref{sec:discussion_benchmark_MC} we benchmark a Monte Carlo
method and its convergence to the exact solution. We find that an accurate description
requires an accounting for dozens of light scatterings. In~\cref{sec:discussion_faster_MC} we also propose a method for faster convergence by changing the parameters of Monte Carlo sampling. Finally, in~\cref{sec:discussion_MC_convergence_rate} we provide a general discussion of the Monte Carlo convergence rate to the true solution.

The technical details of the derivation of the solution and its validation are presented in Appendix~\ref{app:direct_proof}. In Appendix~\ref{app:mc_improve}, we demonstrate that a new Monte Carlo sampling method is equivalent to the original method for an infinite number of scatterings, while converging to the exact solution faster.

Finally, Section~\ref{sec:summary} concludes our study.

\section{Solution to one-dimensional RTE}
\label{sec:the_solution}
Let us denote by $L(l,x,s)$ the absolute value of a flux of photons (units: $\text{m}^{-2}$) at position $x$, time $t$ and direction $s=\pm 1$, corresponding to photon movement to the right ($s=+1$) or to the left ($s=-1$) along $x$. This function obeys the following time-dependent radiative transfer equation, drastically simplified for one-dimension,
\begin{equation}
    \label{eq:rte1}
    \left(\partial_l+ s \partial_x + \mu_t\right) L(l,x,s) = \mu_s \sum_{s'=\pm 1} P(s,s')L(l,x,s'),
\end{equation}
where $\partial_x = \partial/\partial x$, $\partial_l = \partial/\partial l$ with $l=ct$ being the total path length travelled with speed $c$ during time $t$ along the axis $x$, $c$ is the speed of light in the medium,  $\mu_t$ is a sum of absorption $\mu_a$ and scattering $\mu_s$ inverse wavelengths.  The probability density scattering function $P(s,s')$ takes one of two possible values
\begin{equation}
    \label{eq:rte2}
    \begin{aligned}
        P(+1,+1) & =P(-1,-1) = \frac{1+g}{2},\\
        P(+1,-1) & =P(-1,+1) = \frac{1-g}{2},
    \end{aligned}
\end{equation}
with the sum of these quantities equal to one. The parameter $g$ can be interpreted as  {\em scattering asymmetry}: $g=1$ corresponds to the forward scattering, $g=-1$ to the backward scattering, and $g=0$ to equal chances of scattering back and forth.

Introducing
\begin{equation}
    L_\pm(l,x)=e^{\mu_t l}L(l,x,\pm 1),
\end{equation}
~\cref{eq:rte1} could be re-written as a system of coupled equations
\begin{equation}
    \label{eq:rte3}
    \begin{aligned}
        &\partial_l L_++ \partial_x L_+ = \mu_s\left(\frac{1+g}{2}L_++ \frac{1-g}{2}L_-\right),\\
        &\partial_l L_-- \partial_x L_- = \mu_s\left(\frac{1-g}{2}L_++ \frac{1+g}{2}L_-\right).
    \end{aligned}
\end{equation}
As one could observe,~\cref{eq:rte1} does not contain the source function. Instead, we shall consider the initial conditions

\begin{equation}
    \label{eq:rte4}
    \begin{aligned}
        L(0,x,+1) & = L_+(0,x) = \delta(x) ,\\
        L(0,x,-1) & = L_-(0,x) = 0 ,\\
    \end{aligned}
\end{equation}
which correspond to a photon at the origin $x=0$ moving to the right at $t=0$.

Substitutions
\begin{equation}
    \label{eq:rte5}
    \begin{aligned}
        L_+ & = e^{\mu_s (1+g) l/2}\widetilde{L}_+,\\
        L_- & = e^{\mu_s (1+g) l/2}\widetilde{L}_-
    \end{aligned}
\end{equation}
lead to the following system of equations relating $\widetilde{L}_\pm$
\begin{equation}
    \label{eq:rte6}
    \begin{aligned}
        (\partial_l+\partial_x)\widetilde{L}_+ &= \lambda\widetilde{L}_-,\\
        (\partial_l-\partial_x)\widetilde{L}_- &= \lambda\widetilde{L}_+,
    \end{aligned}
\end{equation}
where
\begin{equation}
    \label{eq:rte6a}
    \lambda = \mu'_s/2, \quad \mu_s' = \mu_s(1-g).
\end{equation}
The solution to~\cref{eq:rte6} with initial conditions in~\cref{eq:rte4} is found to be
\begin{equation}
\label{eq:rte7}
  \begin{aligned}
    \widetilde{L}_+(l,x) & = \delta(l-x) + \frac{\lambda}{2}\widetilde{\theta}(\tau^2)\sqrt{\frac{l+x}{l-x}}I_1(\lambda\tau),\\
    \widetilde{L}_-(l,x) & = \frac{\lambda}{2}\widetilde{\theta}(\tau^2)I_0(\lambda\tau),
  \end{aligned}
\end{equation}
with
\begin{equation}
\label{eq:rte8}
\widetilde{\theta}(\tau^2) =(1-\theta(-\tau^2)), \tau = \sqrt{l^2-x^2},
\end{equation}
where $\theta(z)$ and $\widetilde{\theta}(z)$ are two Heaviside functions with a distinct definition at the zero argument
\begin{equation}
\theta(z)=\begin{cases}1,&z\ge 0\\0,&z< 0\end{cases}, \text{ and }
\widetilde{\theta}(z)=\begin{cases}1,&z> 0\\0,&z\le 0\end{cases}.
\end{equation}
The solution to~\cref{eq:rte7} is casual and valid for both domains $l\ge 0$ and $l<0$. Multiplication of $\widetilde{L}_\pm$ by $\theta(l)$ ($\theta(-l)$), also proven  to be the solution to~\cref{eq:rte6}, corresponds to the retarded (advanced) casual Green functions.
A derivation of this solution can be found in~\cref{app:direct_proof}.

Functions $I_n(z)$ in~\cref{eq:rte7} refer to the modified Bessel functions of the first kind. $I_n(z)$ are exponentially growing functions for $z\to\infty$.
However, the solution to~\cref{eq:rte1} is finite everywhere because of the additional exponential factor in~\cref{eq:rte5}
\begin{equation}
    \label{eq:rte9}
    \begin{aligned}
        &L(l,x,+1)  = e^{-(\mu_a+\mu'_s/2) l} \times \\
                  & \times\left(\delta(l-x) + \widetilde{\theta}(\tau^2)\sqrt{\frac{l+x}{l-x}}\frac{\mu'_s}{4}I_1\left(\mu'_s\tau/2\right)\right)\\[0.5pt]
        &L(l,x,-1)  = e^{-(\mu_a+\mu'_s/2) l} \;\widetilde{\theta}(\tau^2)\frac{\mu'_s}{4}I_0\left(\mu'_s\tau/2\right),
    \end{aligned}
\end{equation}
which is the major result of this work.

%\Cref{app:validation} provides a validation of this result by comparing the solution to the numerical Monte Carlo simulation.

\section{Discussion}
\label{sec:discussion}
Let us discuss the obtained solution. (i) For any point $x$ and moment in time $t$ we know the fluxes $L(l,x,\pm 1)$ of photons moving to any direction. (ii) The fluxes are exponentially attenuated by absorption ($\mu_a$) and reduced scattering $\mu_s'=\mu_s(1-g)$ inverse lengths. Let us note that the solution does not depend on $\mu_s$ alone.  (iii) Dirac delta function present in $L(l,x,+1)$ solution corresponds to light keeping its original direction and never experiencing any scattering except on zero angle. The latter observation is supported by the exponential attenuation with reduced scattering $\mu_s'=\mu_s(1-g)$ inverse length rather than by $\mu_s$. (iv) The Modifed Bessel functions in both $L(l,x,\pm 1)$ account for all possible scatterings of light. The argument of the Bessel function is a Lorentz invariant $\sqrt{l^2-x^2}\mu'_s$. A further insight can be gained examing moments of the photons flux.

\subsection{Moments of the flux}
\label{sec:discussion_moments}
Let us study the moments of photons distribution
\begin{equation}
\langle x^n(l)\rangle_s = \frac{1}{N(l)}\int\limits_{-\infty}^{\infty}dx x^n L(l,x,s),
\label{eq:moments1}
\end{equation}
where $L(l,x,s)$ is given by~\cref{eq:rte9}, $s=\pm 1$ and
\begin{equation}
N(l)=\int\limits_{-\infty}^{\infty}dx \left(L(l,x,+1)+L(l,x,-1)\right).
\end{equation}
After some algebra one can get
\begin{equation}
\begin{aligned}
\langle x^{2m}(l)\rangle_+ &=  l^{2m}\frac{\Gamma(m+\frac{1}{2})I_{m-\frac{1}{2}}(\gamma)}{(\gamma/2)^{m-\frac{1}{2}}}e^{- \gamma},\\
\langle x^{2m+1}(l)\rangle_+ &=  l^{2m+1}\frac{\Gamma(m+\frac{3}{2})I_{m+\frac{1}{2}}(\gamma)}{(\gamma/2)^{m+\frac{1}{2}}}e^{- \gamma},\\
\langle x^{2m}(l)\rangle_- &=  l^{2m}\frac{\Gamma(m+\frac{1}{2})I_{m+\frac{1}{2}}(\gamma)}{(\gamma/2)^{m-\frac{1}{2}}}e^{- \gamma},\\
\langle x^{2m+1}(l)\rangle_- &=  0,
\end{aligned}
\label{eq:moments2}
\end{equation}
where
\begin{equation}
\gamma = \mu_s'l/2
\end{equation}
and $\Gamma$ is the Gamma function.

\Cref{eq:moments2} allows us to gain further insight on propagation of light in medium.

(i) The total amount of photons $N(l)$ decreases exponentially
\begin{equation}
N(l)=e^{-\mu_a l}
\end{equation}
due to the absorption.

(ii) Mean coordinate $\langle x(l)\rangle= \langle x(l)\rangle_+ + \langle x(l)\rangle_-$
of photons grows linearly with time until $l\ll {\mu'_s}^{-1}$ and at larger times $l\gg{\mu'_s}^{-1}$
has a well known~\cite{zaccanti1994} limit
\begin{equation}
\begin{aligned}
\langle x(l)\rangle &= l\frac{I_{\frac{1}{2}}(\gamma)}{(\gamma/2)^{\frac{1}{2}}}\Gamma(\frac{3}{2})e^{- \gamma}.\\
\langle x(l)\rangle &\sim \left\{
\begin{array}{ll}
l, &\text{ for } l\ll {\mu'_s}^{-1},\\
{\mu'_s}^{-1}, &\text{ for } l\gg{\mu'_s}^{-1}.
\end{array}
 \right.
\end{aligned}
\end{equation}
When photon's mean coordinate $\langle x(l)\rangle$ reaches the critical value ${(\mu'_s)^{-1}}$,
the mean velocity of the photon's flux vanishes. Any further displacement of photons beyond this limit
is possible only because of diffusion. To see this one have to examine the second order
moment.

(iii) The mean coordinate squared $\langle x^2(l)\rangle= \langle x^2(l)\rangle_+ + \langle x^2(l)\rangle_-$
can be  found with help of~\cref{eq:moments2}
\begin{equation}
\begin{aligned}
\langle x^2(l)\rangle &= l^2\frac{\left(I_{\frac{3}{2}}(\gamma)+I_{\frac{1}{2}}(\gamma)\right)}{(\gamma/2)^{\frac{1}{2}}}\Gamma(\frac{3}{2})e^{- \gamma}.\\
\langle x^2(l)\rangle &\sim\left\{
\begin{array}{ll}
l^2, &\text{ for } l\ll {\mu'_s}^{-1},\\
l{\mu'_s}^{-1}, &\text{ for } l\gg{\mu'_s}^{-1}.
\end{array}
 \right.
\end{aligned}
\end{equation}
One can notice that $\langle x^2(l)\rangle$ grows $\sim l^2$ at small times and keeps growing at smaller rate  $\sim l$  at larger times $l\gg{\mu'_s}^{-1}$ when $\langle x(l)\rangle$ stops growing.  This is why the photon's flux diffuses further in space.

The dispersion $D(l)= \langle x^2(l)\rangle - \langle x(l)\rangle^2$ can be found as follows
\begin{equation}
\begin{aligned}
D(l)  &= l^2\frac{\Gamma(\frac{3}{2})e^{- \gamma}}{(\gamma/2)^{\frac{1}{2}}}\left(
I_{\frac{3}{2}}(\gamma)+I_{\frac{1}{2}}(\gamma)- \frac{\Gamma(\frac{3}{2})I^2_{\frac{1}{2}}(\gamma)}{(\gamma/2)^{\frac{1}{2}}}
\right).\\
D(l) &\sim\left\{
\begin{array}{ll}
0, &\text{ for } l\ll {\mu'_s}^{-1},\\
l{\mu'_s}^{-1}, &\text{ for } l\gg{\mu'_s}^{-1}.
\end{array}
 \right.
\end{aligned}
\end{equation}
Therefore, light keeps expanding further at space beyond ${\mu'_s}^{-1}$ limit due to
diffusion with constant rate $\dot{D}(l)=c{\mu'_s}^{-1}$ at $l\gg{\mu'_s}^{-1}$.

\subsection{Absence of Superluminal Solutions: A Remarkable Result}
\label{sec:tachyon}

A noteworthy observation arises when applying $(\partial_l-\partial_x)$ to the first term and $(\partial_l+\partial_x)$ to the second term of~\cref{eq:rte6}. Both $\widetilde{L}_\pm$ satisfy the Klein-Gordon equation for one spatial dimension, but with an {\em imaginary} ``mass-term'' $m\to i\lambda$:
\begin{equation}
\label{eq:rte10}
\left(\partial^2_l-\partial^2_x-\lambda^2\right)\widetilde{L}_\pm=0.
\end{equation}

At first glance, one might anticipate a {\em superluminal} and exponentially growing solution for this equation, as seen in some examples in~\cite{PhysRev.182.1400}. Such a solution would pose a significant problem for a causal light transport problem.

However, our solution in~\cref{eq:rte9}, when combined with the appropriate boundary conditions, is {\em strictly causal}. Any attempt to replace $\widetilde{\theta}(\tau^2)$ with the acausal $\theta(-\tau^2)$ would violate the boundary conditions. Moreover, the solution $\widetilde{L}_+$ bears resemblance to the retarded Green function of the Klein-Gordon equation with the replacement $m\to i\lambda$, where $\lambda>0$. This replacement transforms the Bessel function $J_1$ into the modified Bessel function $I_1$.

Our findings present an unexpected yet profound illustration: the exact analytic solution in~\cref{eq:rte9} for the light transport problem demonstrates that the Klein-Gordon equation with an {\em imaginary} ``mass-term'' can indeed be causal under appropriate boundary conditions. This result stands in stark contrast to naive expectations and underscores the importance of careful analysis in the study of light transport problems.

\subsection{Benchmarking the Monte Carlo methods}
\label{sec:discussion_benchmark_MC}
Exact solutions provide a benchmark to investigate the applicability range of Monte Carlo methods. It is well known that Monte Carlo methods converge asymptotically to the exact solution when the number of photons ($N_\gamma$) and the number of scattering events ($n_\text{scat}$) are both infinitely large.

To explore this issue, we developed a simple Monte Carlo algorithm that models $n_\text{scat}$ random scattering events of a photon:

\begin{enumerate}
\item The initial conditions are set as follows: the photon is initially located at the origin ($x_0=x(0)=0$) and propagating in the positive direction ($s_0=s(0)=+1$).
\item At each scattering event $i$, the algorithm samples the photon's displacement distance $l_i$ and new propagation direction $s_i$. The displacement distance is sampled according to the probability density function $e^{-\mu_s l}$, and the new direction is sampled according to the probability density function:
\begin{equation}
s_i = \left\{
\begin{array}{ll}
+s_{i-1}, & \text{with probability }(1+g)/2,\\
-s_{i-1}, & \text{with probability }(1-g)/2.
\end{array}
\right.
\end{equation}

Here, $g$ is the anisotropy factor, which characterizes the angular distribution of the scattered photons, and $\mu_s$ is the scattering coefficient.
\item After $n_\text{scat}$ scattering events, the photon's final position is given by
\begin{equation}
x = \sum_{i=0}^{n_\text{scat}} x_i + s_i\Delta x_i,
\end{equation}
where $\Delta x_i = |x_{i+1}-x_i|$ and $x_i$ is the photon's position after the $i$-th scattering event. The total path length $l$ is given by
\begin{equation}
l=\sum_{i=0}^{n_\text{scat}} l_i.
\end{equation}
\item We repeated this procedure $N_\gamma$ times to compute histograms of $L_\pm(l,x)$, where $L_\pm$ are the incoming and outgoing light intensities, respectively. The absorption coefficient $\mu_a$ was included as a multiplication factor.
\end{enumerate}

\begin{figure}[!h]
\centering
\begin{tabular}{c}
\includegraphics[width=\linewidth]{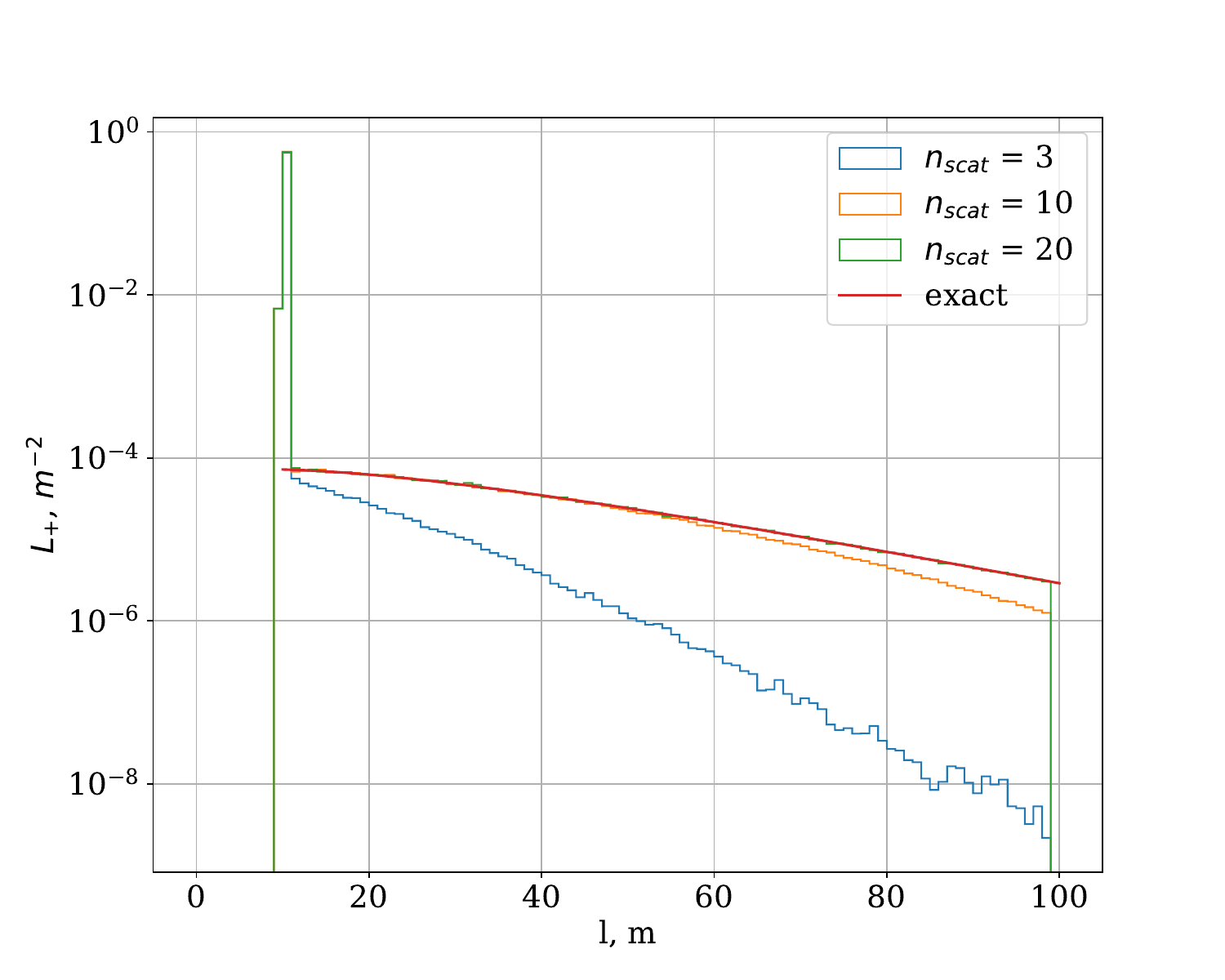}\\
\includegraphics[width=\linewidth]{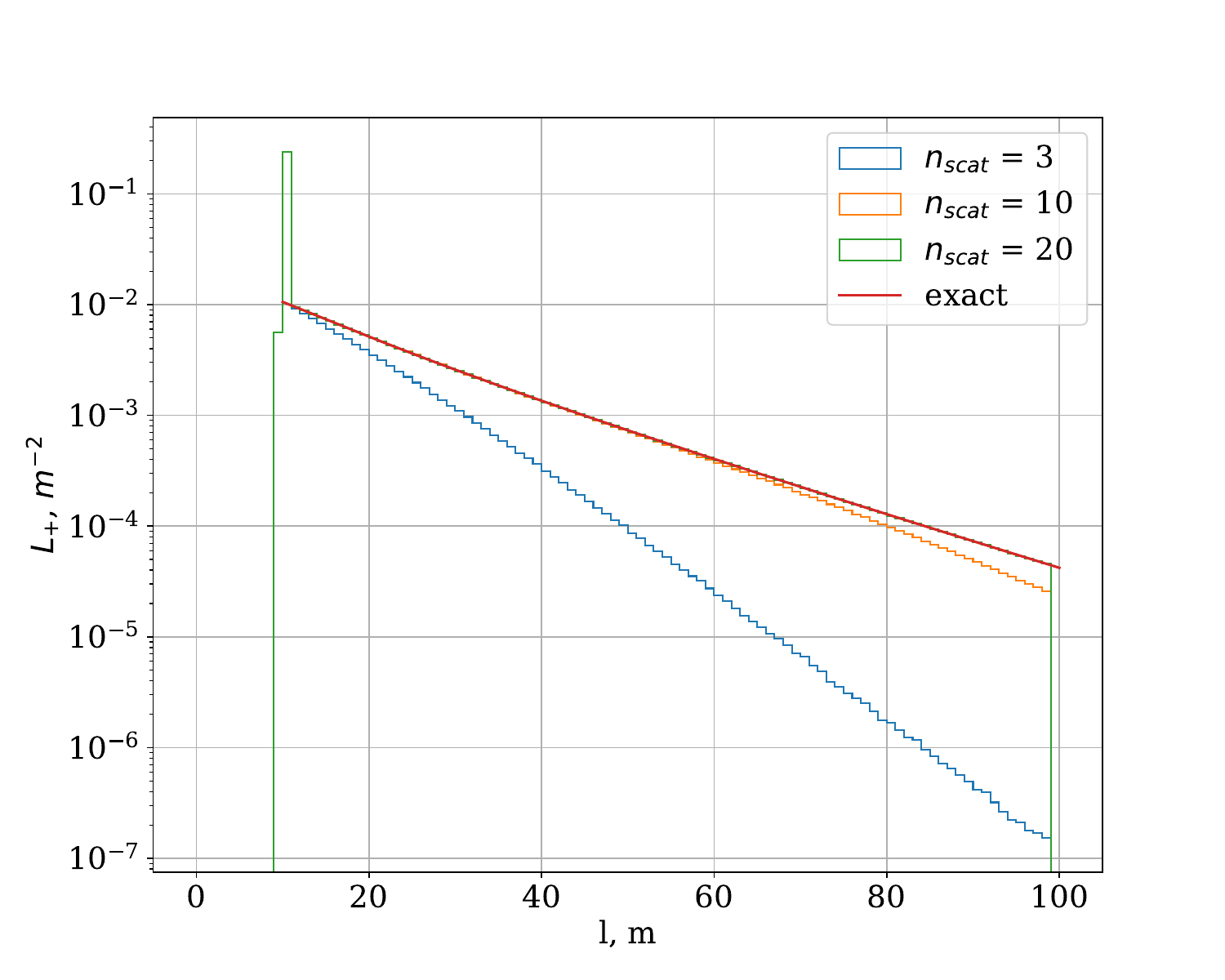}
\end{tabular}
\caption{Comparison of the exact solution $L_+(l)$, where $l=ct$, (solid line) with the Monte Carlo estimates using $N_\gamma=10^7$ and $n_\text{scat}=3,10,20$, assuming $\mu_a=0.05 m^{-1}$, $\mu_s=0.1 m^{-1}$, $x=10\rm{m}$ and $g=0.9$ (upper plot) and $g=-0.9$ (bottom plot).}
\label{fig1}
\end{figure}

In~\cref{fig1}, we compare the exact solutions to the Monte Carlo estimates with $N_\gamma=10^7$ and $n_\text{scat}=3,10,20$ for the anisotropy factor $g=0.9$ and $g=-0.9$, as two extreme examples of a highly anisotropic scattering.
One can conclude, based on examination of~\cref{fig1}, that the Monte Carlo method converges slowly to the exact solution and a dozens of scattering events are required for an accurate estimate of the photon's flux even for highly forward anisotropy.

\subsection{A faster Monte Carlo}
\label{sec:discussion_faster_MC}
An interesting observation is that the exact solution in equation~\eqref{eq:rte9} does not depend on $\mu_s$ and $g$ separately, but rather on the quantity $\mu'_s = (1-g)\mu_s$. One might wonder if this observation can be used to improve the Monte Carlo algorithm presented earlier. Specifically, one could replace $\mu_s$ with $\mu'_s/2$ and $g$ with some other value. But what value should be chosen for $g$? A plausible guess is that $g=-1$ is the appropriate choice.

Physically, this guess is supported by the fact that an effective inverse scattering length of $(1-g)\mu_s$ corresponds to excluding forward scattering, where the direction of a photon remains unchanged. By replacing $\mu_s$ with $\mu'_s/2$, the scattering length becomes longer, but scattering events are still present. In the one-dimensional problem, the only choice available is to change the direction to the opposite direction when such an event occurs. This consideration reinforces the hypothesis that the following replacements are equivalent in the Monte Carlo algorithm:
\begin{equation}
\begin{aligned}
\mu_s &\to \mu'_s/2,\\
g & \to -1.
\end{aligned}
\label{eq:the_substitution}
\end{equation}
In appendix~\ref{app:mc_improve}, we prove this statement.
\begin{figure}[!h]
\centering
\begin{tabular}{c}
\includegraphics[width=\linewidth]{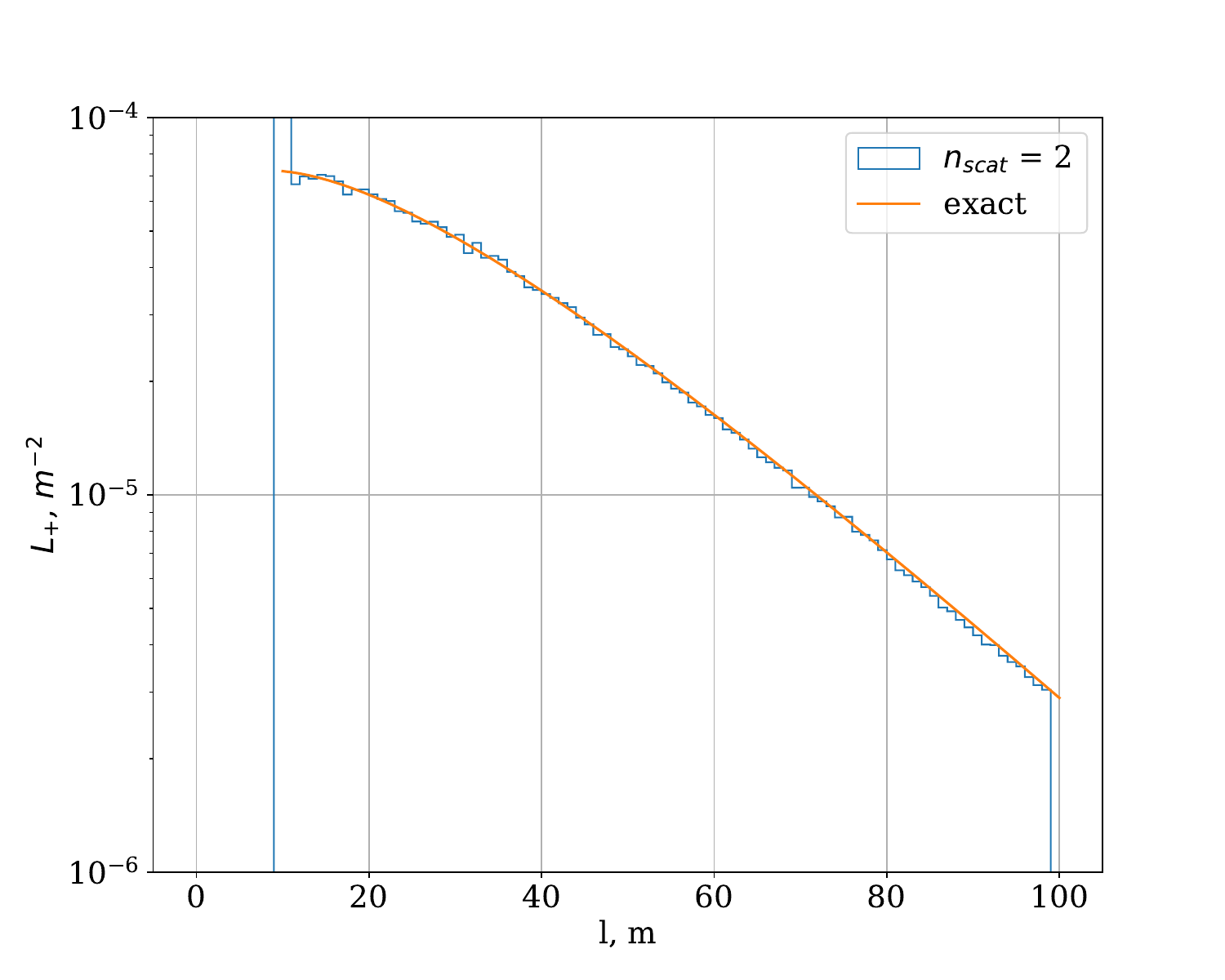}\\
\includegraphics[width=\linewidth]{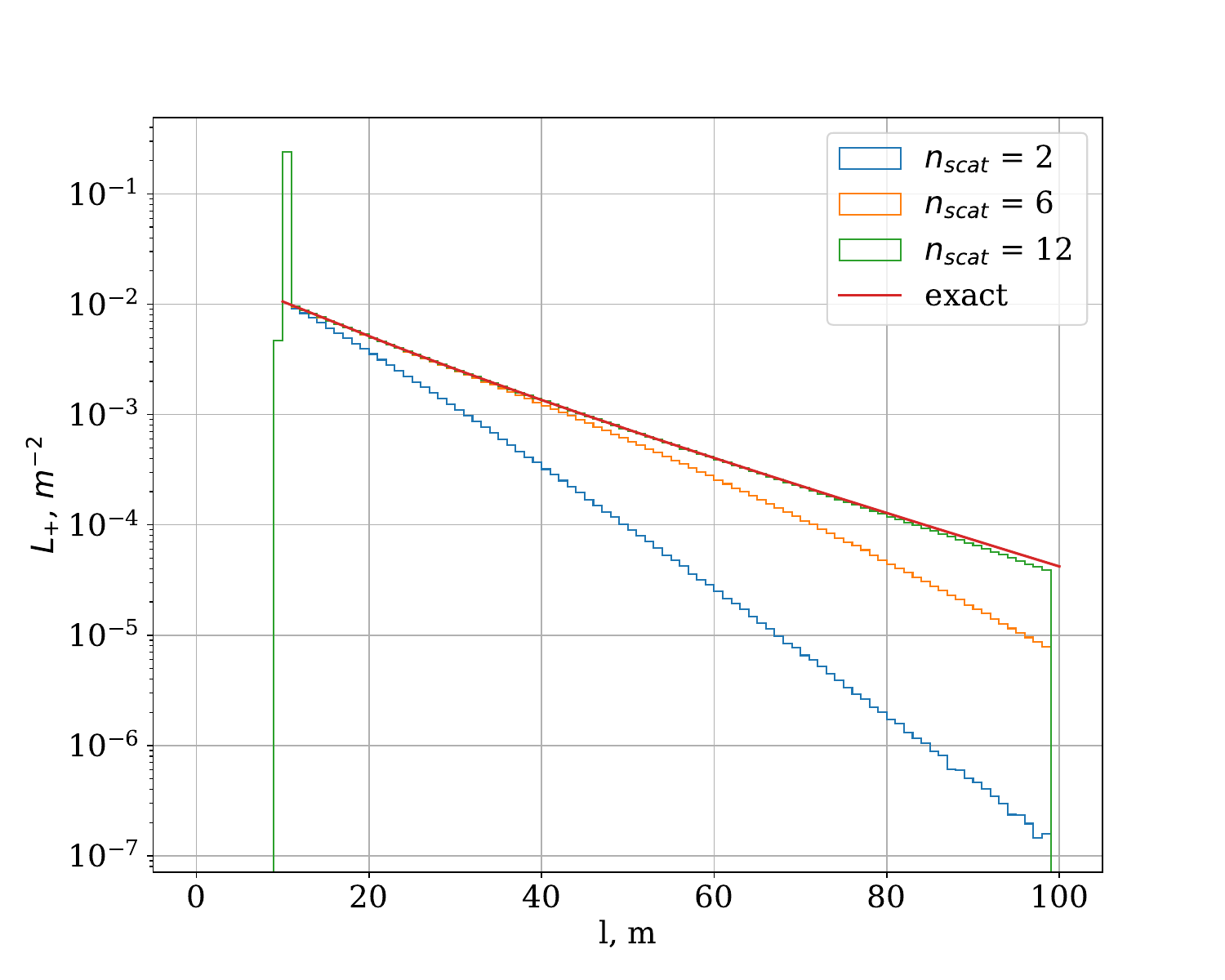}
\end{tabular}
\caption{Comparison of the exact solution $L_+(l)$, where $l=ct$, (solid line) with the improved Monte Carlo estimates (histograms) using $N_\gamma=10^7$ and assuming $g=0.9$ (upper, $n_\text{scat}=2$) and $g=-0.9$ (bottom, $n_\text{scat}=2,6,12$), $\mu_a=0.05 m^{-1}$, $\mu_s=0.1 m^{-1}$ and $x=10\rm{m}$.}
\label{fig2}
\end{figure}
Although the Monte Carlo algorithm provides equivalent results for both input tuples $(\mu_s,g)$ and $(\mu'_s,-1)$, the latter demonstrates a much faster convergence rate, as can be observed in the comparison of the Monte Carlo estimates in~\cref{fig2} and the corresponding figure in~\cref{fig1}. These figures were  generated using the same parameter values for $N_\gamma$, $\mu_a$ and $\mu_s$ and distict asymmetry parameter $g=0.9$ (upper plots) and $g=-0.9$ (bottom plots). A significant increase in the convergence rate can be observed for highly forward scattering ($g=0.9$) where just two scatterings are enough for an accurate estimate of the photon's flux in the given time domain interval (we remind a reader that $l=ct$). However even for highly backward scattering  ($g=-0.9$) the improved version of the Monte Carlo algorithm requires fewer number of scatterings to accurately etimate the photon's flux: twelth (improved) vs twenty (original). At most extreme case $g=-1$ both Monte Carlo version show {\em identical} performance. Therefore, closer asymmetry parameter $g$ approaches to one, fewer scattering
are required for the improved version of Monte Carlo.

\subsection{Convergence Rate of Monte Carlo Method}
\label{sec:discussion_MC_convergence_rate}

Accurately estimating the photon's flux in a Monte Carlo method requires determining the required number of scattering steps $n_\text{scat}$, which can be addressed using the exact solution.

We determine $n_\text{scat}$ as follows:

(i) Fix a specific value of the spatial coordinate $x$ expressed in $\mu_s^{-1}$.

(ii) Considering the time interval where $l-|x|\le k\mu_s^{-1}$, where $k$ is a free parameter, we require that for any $l\in (|x|,|x|+k\mu_s^{-1})$, the following condition holds:
\begin{equation}
| L_\pm^\text{exact}(l,x)-L_\pm^\text{MC}(l,x|n_\text{scat})|<\varepsilon L_\pm^\text{exact}(l,x),
\label{eq:determination_n_scat}
\end{equation}
Here, $L_\pm^\text{exact}(l,x)$ represents the exact solution given by~\cref{eq:rte9}, and $L_\pm^\text{MC}(l,x|n_\text{scat})$ is the corresponding Monte Carlo approximation that depends on the number of scattering steps $n_\text{scat}$. The parameter $\varepsilon$ represents the desired precision. Thus, we estimate $n_\text{scat}$ for both the original and faster Monte Carlo algorithms, resulting in two numbers: $n_\text{scat}^{orig}$ and $n_\text{scat}^{fast}$.

\begin{figure}[!h]
\centering
\includegraphics[width=\linewidth]{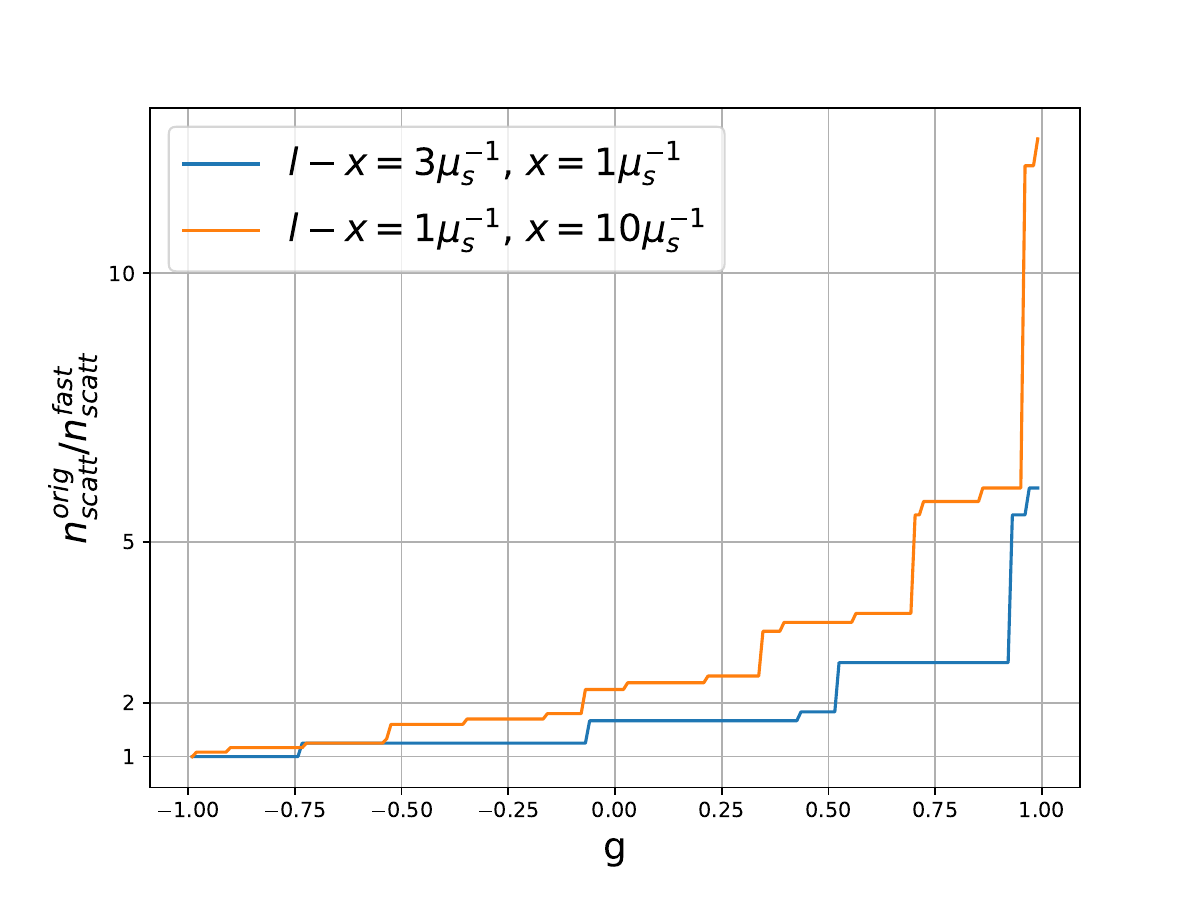}
\caption{The ratio $n_\text{scat}^\text{orig}/n_\text{scat}^\text{fast}$ as a function of the asymmetry parameter $g$ for several values of $k$ and $x$, assuming $\varepsilon=10^{-3}$.}
\label{fig:n_scatterings}
\end{figure}

In~\cref{fig:n_scatterings}, we present the ratio $n_\text{scat}^\text{orig}/n_\text{scat}^\text{fast}$ as a function of the asymmetry parameter $g$ for various values of $k$ and $x$, assuming $\varepsilon=10^{-3}$. This figure illustrates the anticipated trend: a significantly fewer number of steps is required for the faster algorithm as $g\to 1$.

\section{Summary}
\label{sec:summary}
The radiative transfer equation describes how electromagnetic radiation propagates through a medium, and numerical methods are commonly used to approximate its solution. In this paper, we present an exact solution to the one-dimensional time-dependent RTE, which has important practical applications in fields such as atmospheric science, astrophysics, remote sensing, and biomedical optics. The solution is validated using Monte Carlo simulations, and we provide a quantitative analysis of the convergence of the Monte Carlo method to the exact solution.

It is well known that even for highly forward scattering, dozens of random light scatterings are required to obtain an accurate estimate. We demonstrate that a Monte Carlo method with input parameters $(\mu_s(1-g)/2,-1)$ can sample photon scatterings much faster than that with $(\mu_s,g)$ as an input tuple, without sacrificing accuracy. Our study may have implications for other numerical methods used to approximate the RTE.

Interestingly, the exact solution satisfies the Klein-Gordon equation with an imaginary ``mass'' term, which might suggest a superluminal motion. However, the actual solution is found to be strictly causal.

\section{Acknowledgements}
The authors express their gratitude to the Joint Institute for Nuclear Research for providing financial support for this research. We also appreciate the valuable insights and fruitful discussions contributed by Prof. V. A. Naumov throughout the course of this study.

\bibliographystyle{unsrt}
\bibliography{references}  %%% Remove comment to use the external .bib file (using bibtex).
\appendix
\section{Derivation of solution to one-dimensional RTE}
\label{app:direct_proof}
Let us introduce, for the sake of compactness, the differential operators $\hat{K}_\pm =\partial_l\pm\partial_x$.
We begin with a {\em guess} that $\widetilde{L}_-(l,x)$ is actually a function of a single variable $\tau$ defined in~\cref{eq:rte8}. Thus, the second~\cref{eq:rte6} reads as follows
\begin{equation}
    \label{eq:app1}
    \widetilde{L}_+(l,x) = \frac{1}{\lambda}\hat{K}_-\widetilde{L}_-(\tau) = \frac{1}{\lambda}\widetilde{L}'_-(\tau)\frac{l+x}{\tau}.
\end{equation}
As one can see, $\widetilde{L}_+(l,x)$ is not reduced to a function of $\tau$ only.
Substituting~\cref{eq:app1} into the first~\cref{eq:rte6}, one gets
\begin{equation}
    \label{eq:app2}
        \widetilde{L}_-''+\frac{\widetilde{L}'_-}{\tau}-\lambda^2\widetilde{L}_-=0,
\end{equation}
which turns out to be a differential equation with  the single variable, thus verifying the above assumption.

A solution to~\cref{eq:app2} is the modified Bessel function of the first order $I_0(\lambda\tau)$. However, at $l=0$, this solution does not satisfy the initial condition $\widetilde{L}_-(0,x)=0$ since $I_0(\lambda\sqrt{-x^2})\ne 0$. Multiplying $I_0(\lambda\tau)$ by $\widetilde{\theta}(\tau^2)$ turns out also to be a solution to~\cref{eq:app2}, but it satisfies the initial condition for $\widetilde{L}_-$.
Lastly, a solution to~\cref{eq:app2} is determined up to the normalization factor, which we find to be equal to $\lambda/2$:
\begin{equation}
    \label{eq:app3}
    \widetilde{L}_-(\tau) = \frac{\lambda}{2}\widetilde{\theta}(\tau^2)I_0(\lambda\tau).
\end{equation}
The solution for $\widetilde{L}_+$ can be found from the second of~\cref{eq:rte6}
\begin{equation}
    \label{eq:app4}
    \begin{aligned}
    &\widetilde{L}_+  = \frac{1}{\lambda}\hat{K}_-\widetilde{L}_-\\
    & = \frac{1}{2}\left[\left(\hat{K}_-\widetilde{\theta}(\tau^2)\right)I_0(\lambda\tau)+\widetilde{\theta}(\tau^2)\lambda \left(\hat{K}_-\tau\right)I'_0(\lambda\tau)\right]\\
    &=\left[\delta(l-x)+\widetilde{\theta}(\tau^2)\frac{l+x}{\tau}\frac{\lambda}{2}I_1(\lambda\tau)\right],
    \end{aligned}
\end{equation}
where  the following simple relations are handy
\begin{equation}
    \label{eq:app5.1}
    \begin{aligned}
        &\hat{K}_\pm\tau = \frac{l\mp x}{\tau} = \sqrt{\frac{l\mp x}{l\pm x}},\\
        &\hat{K}_\pm\widetilde{\theta}(\tau^2) = 2\delta(l\pm x).
%        &\hat{K}_+\frac{l+x}{\tau} = \frac{1}{\tau}.
    \end{aligned}
\end{equation}
Here we took into account that
\begin{equation}
    \label{eq:app5.2}
    \begin{aligned}
        &\frac{dI_{0}}{dx} = I_1,\\
        &\frac{dI_n}{dx} = I_{n-1} - \frac{n}{x}I_{n}.
%        &\hat{K}_+\frac{l+x}{\tau} = \frac{1}{\tau}.
    \end{aligned}
\end{equation}
Finally, we have to make sure that $\widetilde{L}_+$ in~\cref{eq:app4} satisfies the first. of~\cref{eq:rte6}
\begin{equation}
    \label{eq:app6}
    \begin{aligned}
    & \hat{K}_+\widetilde{L}_+ =\left[\hat{K}_+\delta(l-x)+
    (\hat{K}_+\frac{l+x}{\tau})\widetilde{\theta}(\tau^2)\frac{\lambda}{2}I_1(\lambda\tau)\right.\\
    &\left.+\frac{l+x}{\tau}\frac{\lambda}{2}\left((\hat{K}_+\widetilde{\theta}(\tau^2))I_1(\lambda\tau)+
    \widetilde{\theta}(\tau^2)\lambda(\hat{K}_+\tau)I_1'(\lambda\tau)\right)\right]\\
    &=\lambda \widetilde{L}_-,
    \end{aligned}
\end{equation}
where the following relations are handy
\begin{equation}
    \label{eq:app7}
    \begin{aligned}
        &\hat{K}_\pm\delta(l\mp x) = 0,\\
        &\delta(l\pm x)I_1(\lambda\tau) = 0,\\
        &\hat{K}_\pm\frac{l\pm x}{\tau} = \frac{1}{\tau}.
    \end{aligned}
\end{equation}

\section{Enhancing the Monte Carlo Method}
\label{app:mc_improve}
The Monte Carlo algorithm discussed in~\cref{sec:discussion} corresponds to the following series:

\begin{equation}
\label{eq:MC_orig}
L_\pm(l,x) = e^{-(\mu_a+\mu_s)l}\sum_{n=0}^{\infty}\mu_s^nL^{(n)}_{\pm}(l,x),
\end{equation}
where $L^{(n)}_{\pm}(l,x)$ represents the contribution of $n$ scattering events to the photon flux. The form of~\cref{eq:MC_orig} can be explained as follows: since $l$ is sampled according to the probability density function $e^{-\mu_s l}$, this justifies the corresponding exponential factor in~\cref{eq:MC_orig}. The probability of $n$ independent scattering events is proportional to $\mu_s^n$. Finally, the factor $L^{(n)}_{\pm}(l,x)$ accounts for changes in the photon's direction during the $n$ scattering events.

The exact solution found in~\cref{eq:rte9} suggests that a Monte Carlo algorithm should have a distinct functional dependence:

\begin{equation}
\label{eq:MC_improved}
L_\pm(l,x) = e^{-(\mu_a+\mu'_s/2)l}\sum_{n=0}^{\infty}\left(\frac{\mu'_s}{2}\right)^n\widetilde{L}^{(n)}_{\pm}(l,x),
\end{equation}
since~\cref{eq:rte9} depends on $\mu'_s/2$ rather than on $\mu_s$ and $g$ separately. By expanding the exact solution as a series of $\mu'_s/2$ powers, the coefficient functions $\widetilde{L}^{(n)}_{\pm}$ in~\cref{eq:MC_improved} can be determined. For instance, for the $L_{+}$ function, the coefficients are given by:
\begin{equation}
\label{eq:app9}
\widetilde{L}^{(n)}_{+}(l,x) =
\begin{cases}
\delta(l-x) &\text{for $n = 0$}\\
0 &\text{for odd $n$}\\
\widetilde{\theta}(\tau^2)\sqrt{\frac{l+x}{l-x}}\frac{\tau^{n-1}}{2^n(n/2-1)!(n/2)!} &\text{for even $n$}
\end{cases}
\end{equation}

One can verify that $\widetilde{L}^{(n)}_{\pm}$ does not depend on either $\mu_s$ or $g$.

Let us investigate if this series can be obtained from~\cref{eq:MC_orig} using an appropriate substitution in~\cref{eq:the_substitution}, which was guessed earlier. To demonstrate this, we can start from the exact solution in~\cref{eq:rte9} and rewrite it as:
\begin{equation}
    \label{eq:exact_rewritten1}
    \begin{aligned}
        &L_+  = e^{-(\mu_a+\mu_s) l} e^{+\mu_s l(1+g)/2}\times \\
                  & \times\left(\delta(l-x) + \widetilde{\theta}(\tau^2)\sqrt{\frac{l+x}{l-x}}\frac{\mu'_s}{4}I_1\left(\mu'_s\tau/2\right)\right).
    \end{aligned}
\end{equation}
Let's apply the substitution $g=-1$ in~\cref{eq:exact_rewritten1}. Then,
\begin{equation}
\begin{aligned}
e^{+\mu_s l(1+g)/2} & \to 1,\\
\mu'_s/2 & \to \mu_s
\end{aligned}
\end{equation}
and
\begin{equation}
\label{eq:exact_rewritten2}
\scalebox{0.9}{$\displaystyle
L_+ = e^{-(\mu_a+\mu_s) l} \left(\delta(l-x) + \widetilde{\theta}(\tau^2)\sqrt{\frac{l+x}{l-x}}\frac{\mu_sI_1\left(\mu_s\tau\right)}{2}\right).$}
\end{equation}

Expanding the $I_1$ function in~\cref{eq:exact_rewritten2} as a series, we obtain the exact form of~\cref{eq:MC_improved} with coefficients given by~\cref{eq:app9}. This confirms our guess that the Monte Carlo algorithms with input tuples $(\mu_s,g)$ and $(\mu_s(1-g)/2,-1)$ are identical.

To compare the performance improvement of the Monte Carlo algorithm using this simple substitution in~\cref{eq:the_substitution}, the coefficients $L^{(n)}_{\pm}$ in~\cref{eq:MC_orig} must also be determined for arbitrary $\mu_s$ and $g$. Considering $L^{(n)}_{+}$ for definiteness and assuming $g\ne -1$, we find:
\begin{equation}
\label{eq:MC_orig_terms}
\begin{aligned}
    L^{(0)}_{+} &=\delta(l-x),\\
    L^{(1)}_{+} &=\frac{(1+g)l}{2}\delta(l-x),\\
    L^{(n)}_{+} &= \frac{l^n(1+g)^{n}}{2^{n}n!}\delta(l-x)+ \widetilde{\theta}(\tau^2)(l+x)\times\\
    &\times\frac{1}{4}(1-g)^2l^{n-2}(1+g)^{n-2}\frac{F_1(a_n,b_n;2;4z^2)}{2^{n-2}(n-2)!}
\end{aligned}
\end{equation}
where $z = \frac{(1-g)}{2(1+g)}\sqrt{1-\frac{x^2}{l^2}}$ and $F_1(a_n,b_n,c,x)$ is the hypergeometric function. The parameters $a_n$ and $b_n$ are defined as follows.
\begin{equation}
a_n = \frac{5-2n+(-1)^n}{4}, \quad b_n = \frac{5-2n+(-1)^{n+1}}{4}.
\end{equation}
\begin{figure}[!h]
\centering
\includegraphics[width=\linewidth]{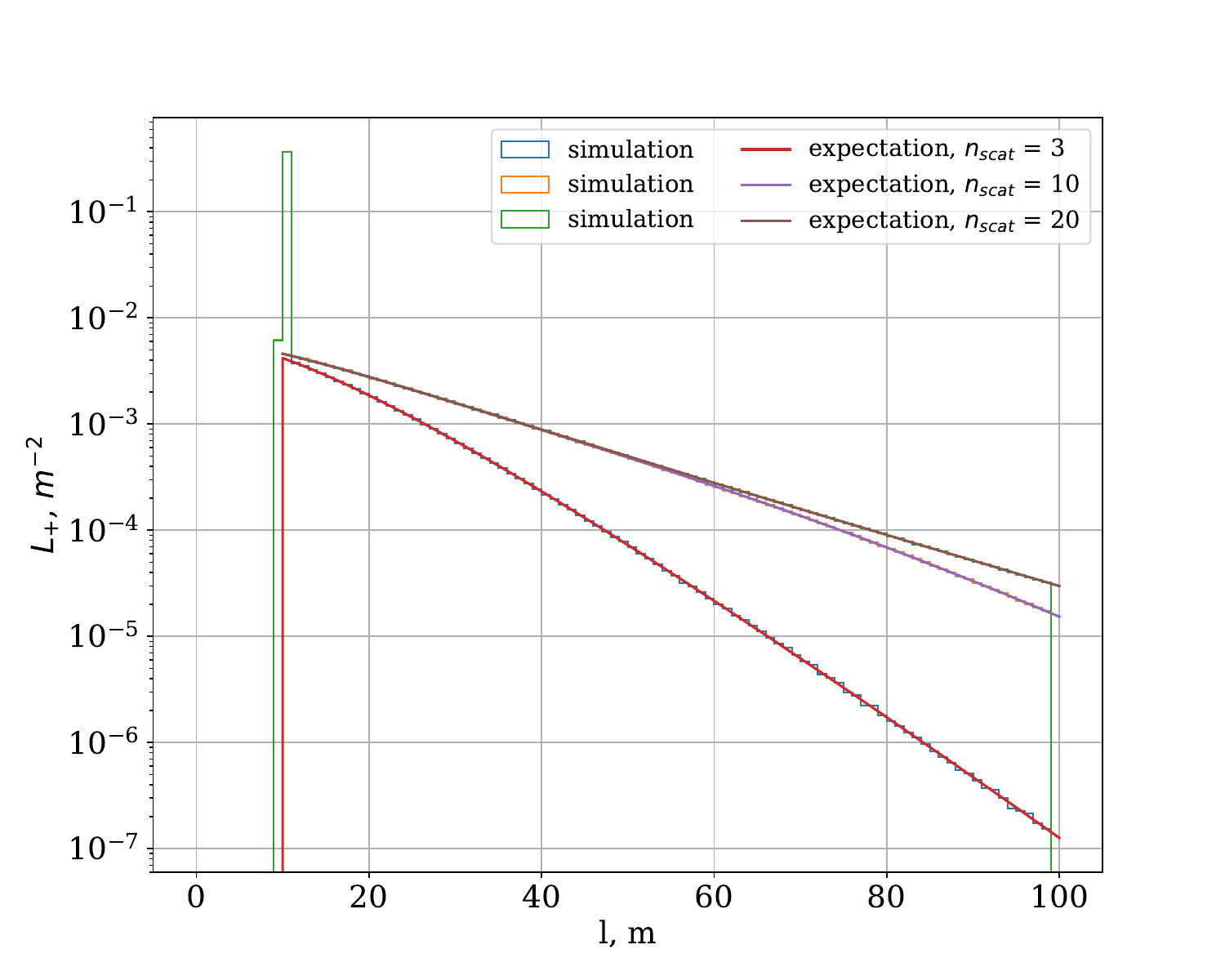}
\caption{Comparison between the analytic expansion of the exact solution $L_+(l)$, where $l=ct$, and the numerical result of Monte-Carlo for $x = 10m$ with $g = 0, \mu_a = 0.05m^{-1}, \mu_s = 0.1m^{-1}$.}
\label{fig:compare_old_mc}
\end{figure}
In~\cref{fig:compare_old_mc}, we compare the results for $L_+$ estimated with the Monte Carlo algorithm discussed in~\cref{sec:discussion} to~\cref{eq:MC_orig}, with $L^{(n)}_{+}$ coefficient functions given by~\cref{eq:MC_orig_terms} for several orders, assuming $g=0$.
All terms up to $n$ are summed up in~\cref{eq:MC_orig} for this comparison. Excellent agreement can be observed, confirming the correctness of~\cref{eq:MC_orig_terms}. One can note that about $n_\text{scat}=20$ terms are required to achieve an accurate estimate.

\begin{figure}[!h]
\centering
\includegraphics[width=\linewidth]{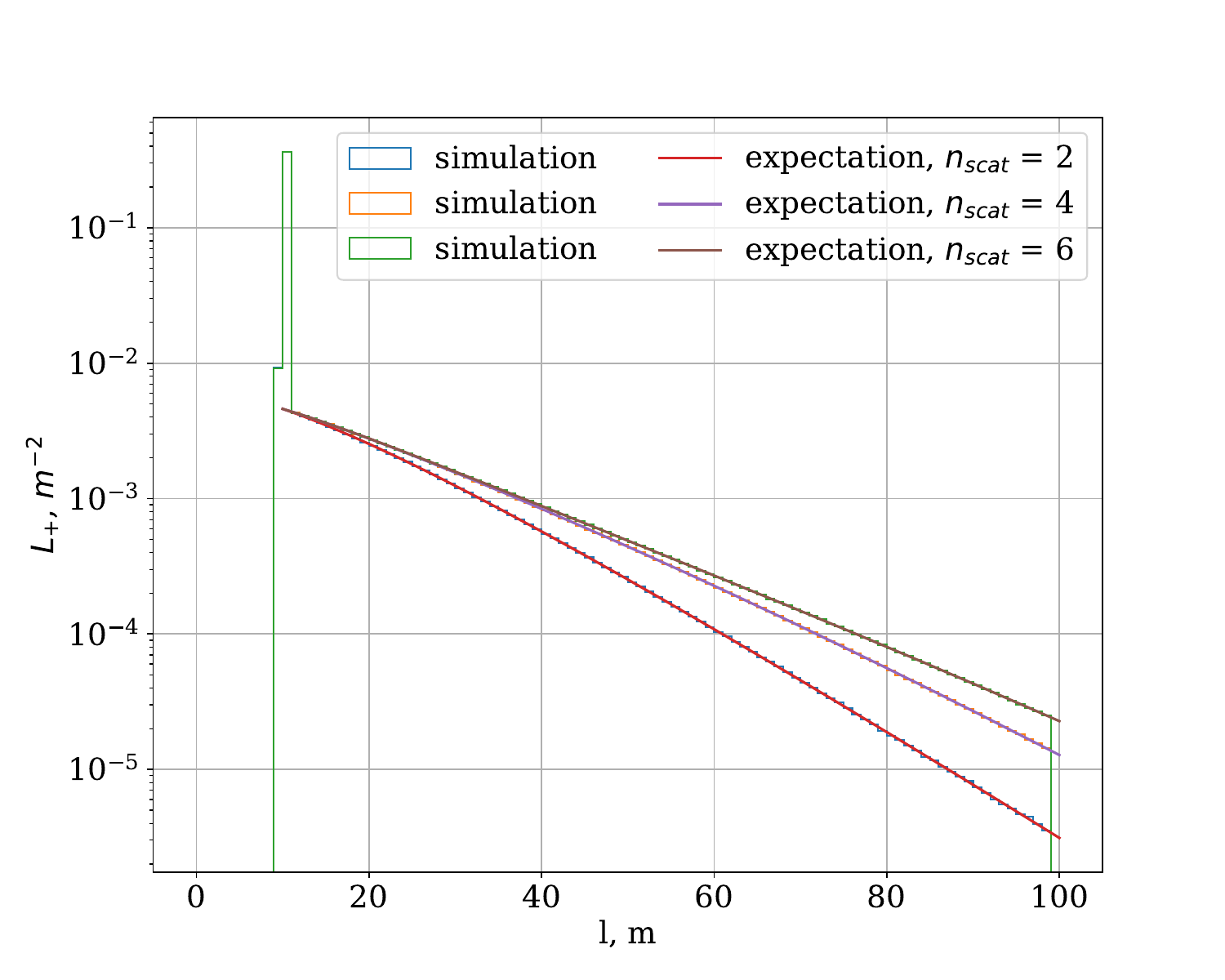}
\caption{Comparison between the analytic expansion of the exact solution $L_+(l)$, where $l=ct$, and the numerical result of improved Monte-Carlo for $x = 10m$ with $g = 0, \mu_a = 0.05m^{-1}, \mu_s = 0.1m^{-1}$.}
\label{fig:compare_new_mc}
\end{figure}

In~\cref{fig:compare_new_mc}, we display a similar comparison for the improved Monte Carlo as given by~\cref{eq:MC_improved,eq:app9}, assuming the same set of parameters.

We can conclude that there is excellent agreement between the analytic coefficients in~\cref{eq:app9} and Monte Carlo expectations, confirming~\cref{eq:MC_improved,eq:app9}. Furthermore, we can observe a much faster convergence to the exact solution with the improved Monte Carlo algorithm.

\end{document}